\documentclass[preprint]{ptephy_v1}
\preprintnumber{LIGO-P2100286}
\usepackage{xcolor}
\usepackage{graphicx}
\usepackage[colorlinks=true,linkcolor=blue,citecolor=blue,allcolors=blue]{hyperref}
\usepackage[caption=false]{subfig}
\usepackage{xspace}
\usepackage[acronym,shortcuts]{glossaries} 
\usepackage{threeparttable}
\usepackage{amsmath}
\usepackage{etoolbox}
\usepackage{orcidlink}
\usepackage{lineno}  
\usepackage{url}
\newcommand{\review}[1]{{\color{red}{#1}}}
\renewcommand{\review}[1]{#1}

\glsdisablehyper

\newacronym{GW}{GW}{gravitational wave}
\newacronym{CBC}{CBC}{compact binary coalescence}
\newacronym{BH}{BH}{black hole}
\newacronym{NS}{NS}{neutron star\glsunset{NSBH}}
\newacronym{BNS}{BNS}{binary \ac{NS}}
\newacronym{NSBH}{NSBH}{\ac{NS}--\ac{BH}}
\newacronym{O1}{O1}{the first observing run of Advanced LIGO and Advanced Virgo}
\newacronym{O2}{O2}{the second observing run of Advanced LIGO and Advanced Virgo}
\newacronym{O3}{O3}{the third observing run of Advanced LIGO and Advanced Virgo}
\newacronym{SNR}{SNR}{signal-to-noise ratio}

\newcommand{\gstlal}{\texttt{GstLAL}\xspace}
\newcommand{\LALSuite}{\texttt{LALSuite}\xspace}
\newcommand{\pygrb}{\texttt{PyGRB}\xspace}
\newcommand{\pycbc}{\texttt{PyCBC}\xspace}
\newcommand{\Xpipeline}{\texttt{X-Pipeline}\xspace}
\newcommand{\xpipeline}{\texttt{X-Pipeline}\xspace}
\newcommand{\omicron}{\texttt{Omicron}\xspace}
\newcommand{\Msun}{\ensuremath{M_{\odot}}\xspace}

\newacronym{CWB}{\texttt{cWB}}{\texttt{coherent WaveBurst}\xspace}
\newacronym{FAR}{FAR}{false alarm rate}
\newacronym{iFAR}{iFAR}{inverse false alarm rate}
\newacronym{SG}{SG}{sine--Gaussian wavelets}
\newacronym{GA}{GA}{Gaussian pulses}
\newacronym{WNB}{WNB}{white noise bursts}
\newacronym{DPF}{DPF}{Dominant Polarization Frame}

\newacronym{ADI}{ADI}{accretion disk instability}
\newacronym{BAT}{BAT}{Burst Alert Telescope}
\newacronym{CSG}{CSG}{circular sine-Gaussian}
\newacronym{GBM}{GBM}{Gamma-ray Burst Monitor}
\newacronym{GCN}{GCN}{Gamma-ray Coordinates Network}
\newacronym{GRB}{GRB}{gamma-ray burst}
\newacronym{VALID}{VALID}{Vetting Automation and Literature Informed Database}

\newcommand{\nCoincGRB}{\review{4}\xspace}

\newcommand{\pvalPYGRBAprFifteen}{\review{0.43}\xspace}
\newcommand{\DBNSAprFifteen}{\review{0.91\,Mpc}\xspace}
\newcommand{\pcBNSAprFifteenNGC}{\review{0\%}\xspace}
\newcommand{\DNSBHGenAprFifteen}{\review{1.08\,Mpc}\xspace}
\newcommand{\pcNSBHGenAprFifteenNGC}{\review{2\%}\xspace}
\newcommand{\DNSBHAliAprFifteen}{\review{1.45\,Mpc}\xspace}
\newcommand{\pcNSBHAliAprFifteenNGC}{\review{9\%}\xspace}

\newcommand{\pvalPYGRBAprTwenty}{\review{0.45}\xspace}
\newcommand{\DBNSAprTwenty}{\review{0.15\,Mpc}\xspace}
\newcommand{\DNSBHGenAprTwenty}{\review{0.21\,Mpc}\xspace}
\newcommand{\DNSBHAliAprTwenty}{\review{0.17\,Mpc}\xspace}

\newcommand{\pvalBurstLowest}{\review{0.132}\xspace}
\newcommand{\nameBurstLowest}{\review{GRB\,200412A}\xspace}

\usepackage{blindtext}

\AtBeginDocument{
  \label{CorrectFirstPageLabel}
  
}

\begin{document}

\title{First joint observation by the underground gravitational-wave detector, KAGRA, with GEO\,600}

\setlength{\textwidth}{7in}
\setlength{\textheight}{9.75in}
\setlength{\topmargin}{-1in}
\setlength{\oddsidemargin}{-.5in}
\setlength{\evensidemargin}{\evensidemargin}
\author{
R.~Abbott$^{1}$,   
H.~Abe$^{2}$,   
F.~Acernese$^{3,4}$,   
K.~Ackley\,\orcidlink{0000-0002-8648-0767}$^{5}$,   
N.~Adhikari\,\orcidlink{0000-0002-4559-8427}$^{6}$,   
R.~X.~Adhikari\,\orcidlink{0000-0002-5731-5076}$^{1}$,   
V.~K.~Adkins$^{7}$,   
V.~B.~Adya$^{8}$,   
C.~Affeldt$^{9,10}$,   
D.~Agarwal$^{11}$
% ...
(The LIGO Scientific Collaboration, the Virgo Collaboration, and the KAGRA Collaboration)
\\
{${}^{\ast}$Deceased, August 2020. }
{${}^{\dag}$Deceased, April 2021. }
}\affil{
$^{1}$LIGO Laboratory, California Institute of Technology, Pasadena, CA 91125, USA 
\\
$^{2}$Graduate School of Science, Tokyo Institute of Technology, Meguro-ku, Tokyo 152-8551, Japan   
\\
$^{3}$Dipartimento di Farmacia, Universit\`a di Salerno, I-84084 Fisciano, Salerno, Italy   
\\
$^{4}$INFN, Sezione di Napoli, Complesso Universitario di Monte S. Angelo, I-80126 Napoli, Italy   
\\
$^{5}$OzGrav, School of Physics \& Astronomy, Monash University, Clayton 3800, Victoria, Australia 
\\
$^{6}$University of Wisconsin-Milwaukee, Milwaukee, WI 53201, USA 
\\
$^{7}$Louisiana State University, Baton Rouge, LA 70803, USA 
\\
$^{8}$OzGrav, Australian National University, Canberra, Australian Capital Territory 0200, Australia 
\\
$^{9}$Max Planck Institute for Gravitational Physics (Albert Einstein Institute), D-30167 Hannover, Germany 
\\
$^{10}$Leibniz Universit\"at Hannover, D-30167 Hannover, Germany 
\\
$^{11}$Inter-University Centre for Astronomy and Astrophysics, Pune 411007, India 
\\
% ...
}

\begin{abstract}
We report the results of the first joint observation of the KAGRA detector with GEO\,600. 
KAGRA is a cryogenic and underground gravitational-wave detector consisting of a laser 
interferometer with three-kilometer arms, and located in Kamioka, Gifu, Japan. 
GEO\,600 is a British--German laser interferometer with 600\,m arms, and located near Hannover, Germany.  
GEO\,600 and KAGRA performed a joint observing run from April 7 to 20, 2020. 
We present the results of the joint analysis of the GEO--KAGRA data for transient 
gravitational-wave signals, including the coalescence of neutron-star binaries and generic 
unmodeled transients. We also perform dedicated searches for binary coalescence signals 
and generic transients associated with gamma-ray burst events observed during the joint run. 
No gravitational-wave events were identified.
We evaluate the minimum detectable amplitude for various types of transient signals 
and the spacetime volume for which the network is sensitive to binary neutron-star coalescences. 
We also place lower limits on the distances to the gamma-ray bursts analysed based on the non-detection 
of an associated gravitational-wave signal for several signal models, including binary coalescences. 
These analyses demonstrate the feasibility and utility of KAGRA as a member of the 
global gravitational-wave detector network.
\end{abstract}

\subjectindex{F31, F32, F33, F34}

\maketitle


\section{Introduction}

The first direct observation of \acp{GW} \cite{GW150914} opened a new branch of astronomy. 
In their first three observing runs, Advanced LIGO and Advanced Virgo have identified 90 candidates 
with probability of astrophysical origin greater than 50\%
\cite{LIGOScientific:2018mvr,LIGOScientific:2020ibl,LIGOScientific:2021usb,LIGOScientific:2021djp},
all of which were consistent with being produced by the inspiral and merger of compact-object binaries comprised 
of \acp{BH} or \acp{NS}.
During the most recent observing run, signals were detected at a rate of greater than 1 event per week 
\cite{LIGOScientific:2020ibl,LIGOScientific:2021djp}, and this rate is expected to grow rapidly as detector
 sensitivity improves \cite{KAGRA:2013pob}.
There is also the potential to detect \acp{GW} from other sources, such as 
core-collapse supernovae \cite{2017hsn..book.1671K,Ott_2009} 
cosmic strings 
\cite{Vachaspati:1984gt,Sakellariadou:1990ne},  
and long \acp{GRB}  \cite{Fryer:2001zw,vanPutten:2003hd,Piro:2006ja,Corsi:2009jt}, 
which would provide probes into the astrophysics of these objects \cite{Sathyaprakash:2009xs,LIGOScientific:2021psn} and further insights into fundamental physics \cite{Sathyaprakash:2009xs,LIGOScientific:2020tif,LIGOScientific:2021sio}.

Optimal use of \ac{GW} data relies on observations by a network of detectors. 
Laser interferometer \ac{GW} detectors are essentially all-sky monitors but have low sky-localization accuracy for short-duration transients.
Determining the source position or host galaxy for short transients relies mostly on triangulation between widely separated detectors \cite{Fairhurst:2009tc,Fairhurst:2010is,Wen:2010cr,Fairhurst:2017mvj,KAGRA:2013pob,Pankow:2019oxl}. 
Multiple detectors with different orientations are also required to disentangle the two wave polarizations, 
which in turn is required, for example, for some tests of general relativity \cite{LIGOScientific:2017ycc,GW150914,LIGOScientific:2018dkp,LIGOScientific:2019fpa,LIGOScientific:2020tif}. 
Measuring both polarizations is also required for determining the source orientation, which is 
needed to determine the distance to binary sources (vital for measurements of the Hubble constant \cite{LIGOScientific:2017adf,DES:2019ccw,LIGOScientific:2019zcs,LIGOScientific:2021bbr}). 
It can also give information on \ac{GRB} beaming \cite{LIGOScientific:2017zic}.
Multiple detectors also provide redundancy against detector downtime and improve the sky coverage of the network.

In this paper we report the results of the first 
joint observation of a new detector in the global network: KAGRA. 
The KAGRA detector \cite{KAGRA:2020tym} took scientific data from April 7 through April 20, 2020, 
at the end of the third observing run (O3) of the LIGO--Virgo--GEO network.
The LIGO and Virgo detectors were forced to terminate operations prematurely due to the COVID-19 pandemic, 
but the GEO\,600 (abbreviated in this paper as GEO) detector continued operations and collected data jointly with KAGRA over this period. 
We present the results of analyses of this joint GEO--KAGRA run data for transient \ac{GW} signals. 
We perform four of the searches that are standard for LIGO--Virgo observing runs. 
Two of these scan all of the data for signals arriving from any direction at any time:  
a search for \ac{BNS} coalescences~\cite{TheLIGOScientific:2017qsa,LIGOScientific:2018mvr,LIGOScientific:2020aai,LIGOScientific:2020ibl}, 
and a search for generic unmodeled short transients (bursts) \cite{LIGOScientific:2016kum,LIGOScientific:2019ppi,LIGOScientific:2021hoh}. 
The other two analyses are dedicated searches for binary coalescence signals and GW bursts associated with \ac{GRB} events observed during the joint run \cite{LIGOScientific:2016akj,LIGOScientific:2019obb,LIGOScientific:2020lst,LIGOScientific:2021iyk}. 
No significant candidate \ac{GW} events are identified, which is expected given the sensitivity of 
KAGRA at this early stage in its commissioning. 
However, the sensitivity of KAGRA is expected to improve by more than two orders of magnitude over the coming years as its design sensitivity is achieved \cite{KAGRA:2013pob}.
These analyses demonstrate the value KAGRA will have as a member of the global network as its sensitivity increases.

This paper is structured as follows. In Section~\ref{sec:run} we describe the KAGRA and GEO detectors, and the joint observing run. 
In Section~\ref{sec:gstlal} we present the all-sky search for \ac{BNS} coalescences.
In Section~\ref{sec:cwb} we present the all-sky search for generic bursts.
In Section~\ref{sec:grb} we present the compact binary coalescence (CBC) and burst searches following up \acp{GRB} observed during the joint run.
We conclude with a  discussion of the prospects for future joint observations in Section~\ref{sec:conclusions}.

\section{GEO--KAGRA Observing Run}
\label{sec:run}

\subsection{KAGRA}

KAGRA \cite{Somiya:2011np,Aso:2013eba,KAGRA:2020tym} is a laser interferometer \ac{GW} 
detector with 3\,km arms, located in Kamioka, Gifu, Japan. 
KAGRA is built underground, and uses cryogenic mirrors for four test masses in two arms.  
Those features help to reduce seismic and thermal noise. 
KAGRA uses sapphire test masses whose diameter, thickness and mass are 22\,cm, 15\,cm and 22.8\,kg, respectively. 

The construction of KAGRA started in 2010. 
However, the start of tunnel excavation was delayed until 2012 due to a major earthquake on March 11, 2011. 
The tunnel excavation was completed by May 2014, then the installation of the laser interferometer started \cite{KAGRA2018,KAGRA:2020tym}. 
The initial test of KAGRA with room temperature mirrors was completed by March 2016, 
and the first operation of the 3\,km Michelson interferometer was done from March to April 2016 \cite{KAGRA2018,KAGRA:2020tym}. 
After the cryogenic systems and mirrors were installed, a test operation of the interferometer 
with one cryogenic mirror was performed from April 28 to May 6, 2018 \cite{KAGRA:2019htd}.

By April 2019, 
most of the interferometer components had been installed, and the commissioning work started. 
In August 2019, the first lock of the Fabry--Perot Michelson interferometer configuration was achieved. 
The first lock of the power-recycled Fabry--Perot Michelson interferometer (PRFPMI) configuration was accomplished in January 2020. 
The signal readout scheme was upgraded from a conventional radio-frequency (RF) readout to a direct-current (DC) readout with an output mode cleaner in February 2020. 
The injected laser power was 5\,W.
The power recycling gain for the carrier field in the PRFPMI configuration was measured to be 
around 11--12. 
The circulating power in the Fabry--Perot arm cavities was 21--25\,kW per arm.

Over the course of six months from August 2019, the detector noise floor was reduced by 3--4 orders of magnitude. 
A standard measure of interferometer sensitivity is the volume- and angle-averaged distance to which the 
inspiral of a 1.4\,$M_\odot$--1.4\,$M_\odot$ binary system can be detected with a matched-filter \ac{SNR} of at least 8 \cite{Finn:1992xs,KAGRA:2013pob}.
From February 25 to March 10, 2020, KAGRA conducted observations with a \ac{BNS} 
observable range of about 600\,kpc.
After further commissioning work, the sensitivity of KAGRA was improved to reach a \ac{BNS} observable range of approximately 1\,Mpc 
by the end of March. 
KAGRA then performed an observation run jointly with GEO from April 7 through 20, 2020.
Since the thermal noise was not a major noise source at this point, the test-mass mirrors were not cooled during this run. 
Further details of the detector design and construction history are given in \cite{KAGRA:2020tym}.

The sensitivity of KAGRA during the joint GEO--KAGRA run was limited at low frequencies 
(below 100\,Hz) by the local control noise of the mirror suspensions, arising from insufficiently 
optimised damping control filters. Above 400\,Hz, the sensitivity was limited by laser shot noise.
At intermediate frequencies the noise is not well-modelled but shows some coherence with 
environmental acoustic noise, which may arise from scattered light coupling.

During the joint run, data is flagged as being in \textit{observing mode} when the PRFPMI configuration is locked with DC readout. 
Fixed-frequency lines are added to the test-mass feedback control signals to calibrate the data. 
The feedback control signals are monitored for saturations or other anomalies, and the data acquisition 
system is checked offline for errors. 
If any anomalies are found in these checks, the observing mode flag is removed. 
The \ac{GW} searches presented in this paper are performed exclusively on data that are flagged 
as observing mode, except for the analysis of GRB\,200415A in Section~\ref{sec:grb}. 
At the time of GRB\,200415A, the detector was locked, but there were a few personnel still near the detector  
following earlier maintenance work.  Thus, the data at this time was not flagged as observing mode. 
However, subsequent investigation of the data found no anomalies, and we 
conclude that we can use the data around the time of GRB\,200415A for \ac{GW} searches. 

\subsection{GEO\,600}

GEO \cite{Luck:2010rt,Affeldt2014Advanced600,Dooley2016GEOChallenges} is a 
British--German interferometric \ac{GW}
detector with 600\,m arms located near Hannover, Germany.
Similar to other \ac{GW} detectors, the design is based on a Michelson interferometer
with a number of features to enhance the sensitivity.
The \ac{GW} signal is read out by controlling the differential arm length 
slightly off of the dark fringe in order to couple the differential arm motion
to the direct-current power at the output.
At high frequencies, the detector is limited by quantum shot noise.
The shot noise originates as vacuum fluctuations entering the interferometer at the output.
By replacing the normal vacuum fluctations with a squeezed vacuum, the quantum noise is
reduced in the measurement quadrature \cite{PhysRevLett.126.041102}.

In contrast to the KAGRA detector, the test masses of GEO are made of fused silica
and operate at room temperature \cite{Winkler2007TheOptics}.
Their diameter, thickness and mass are 18\,cm, 10\,cm and 5.6\,kg, respectively.
The power injected is about 3 W, which leads to about 3 kW of circulating power in the power
recycling cavity which is then 1.5 kW circulating power per arm. 
GEO uses folding in the arms to give an optical length of 1200\,m for each
arm \cite{Affeldt2014Advanced600,Dooley2016GEOChallenges}. 

Normally, the GEO detector is operated in data-taking \textit{astrowatch} mode when the detector
is not being used for instrument science research.
For the joint GEO--KAGRA run period, the detector was operated in a stable 
configuration that included squeezed vacuum injection for increased sensitivity.
The squeezer has a high duty cycle; squeezing was applied for 97.9\% of the observation time.

\subsection{Joint Observing Run and Data Quality}

\begin{figure}
\includegraphics[width=0.46\textwidth]{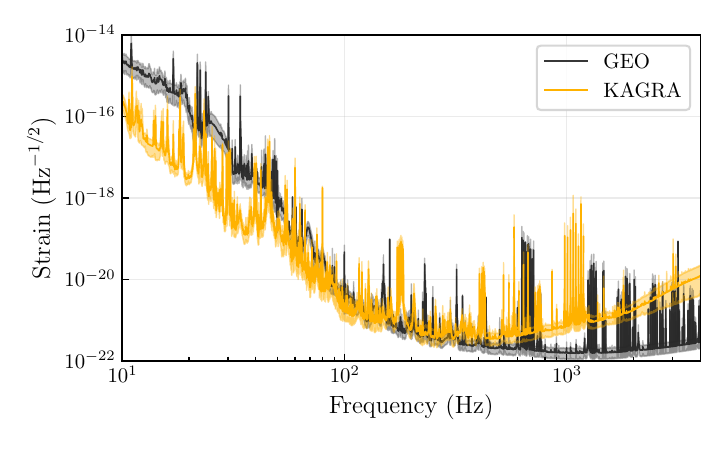}
\includegraphics[width=0.46\textwidth]{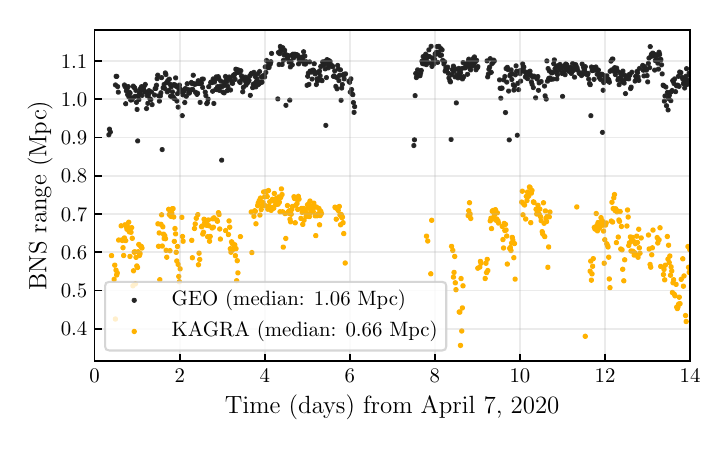}
\caption{Left: Noise amplitude spectral density of GEO (black) and KAGRA (yellow) during the joint observing run. 
The solid curves show the mean sensitivity for each frequency bin and the shaded regions 
show the 5th and 95th percentile over the period.
Narrow peaks in the spectra are due to such sources as resonances of the suspension system (violin modes) and harmonics of the electrical grid frequency (50\,Hz for GEO and 60\,Hz for KAGRA) \cite{openDataG,openDataK}.
Right: \ac{BNS} inspiral ranges for GEO and KAGRA over the joint run.   
The gap around day 6 and 7 was caused when both detectors were affected by bad weather 
and were unable to lock.
\label{fig:spectra}}
\end{figure}

The GEO--KAGRA joint run period was between April 7 2020 08:00 UTC and 
April 21 2020 00:00 UTC.
Figure~\ref{fig:spectra} shows representative sensitivities of the detectors during the run, 
as measured by the amplitude spectral density of the calibrated strain output, 
and the evolution of the detectors' sensitivity over time, as measured by the \ac{BNS} inspiral range. 

Table \ref{tab:obstime} shows the observing times and the duty cycles for two interferometers, 
the latter defined as the percentage of the total run duration in which the instruments were observing. 
The duty cycle of KAGRA was lower than that of GEO for several reasons. 
One was that alignment sensing and control using wavefront sensors was not implemented 
by the time of the run, so that the interferometer could not be operated for long periods. 
Furthermore, following loss of lock of the interferometer it often took a long time to adjust the alignment 
in order to recover lock. 

\begin{table}[htbp]
\centering
\begin{tabular}{l c c }
\hline
& Observing time (days) & Duty cycle \\
\hline 
GEO & 10.90  & 79.8\% \\ 
KAGRA & 7.29 & 53.3\% \\
coincident &  6.39 & 46.8\% \\ 
\hline
\end{tabular}
\caption{The time length of the observing mode and the duty cycle for GEO and KAGRA for the 
period April 7 2020 08:00 UTC to April 21 2020 00:00 UTC.}
\label{tab:obstime}
\end{table}

While the quiet underground environment of KAGRA provides advantages in the 
operation of the instrument, KAGRA is not completely free from the effects of bad weather. 
The nearest coastline is approximately 40\,km away. Ocean waves crashing on the shoreline 
constantly excite ground vibrations around $\sim$0.2\,Hz, which become about one order of magnitude 
stronger during storms. 
The gap between day 6 and 7 in the \ac{BNS} range time series data shown in Fig.~\ref{fig:spectra}
is a period when KAGRA could not operate due to a storm caused by a low-pressure system 
that passed through Japan at that time.

Following the joint run the vibration isolation control system has been improved and 
additional environmental monitors and the wavefront sensor system have been installed.
This has led to an increase in KAGRA's duty cycle. 

The strain data from each interferometer is generated by processing and combining raw electronic signals coming from the
differential arm length control using a detailed model of the control system including the optical
response of the interferometer. Any errors in the measurements which inform the model will lead to
a systematic error in the calibration. In general, systematic error is complex-valued,
frequency-dependent, and time-dependent.
The calibration uncertainty of the data used in this paper are within $\pm 10$\% in amplitude and within $\pm 10$\,$\deg$ in phase (68\% C.L.) between 30\,Hz and 1500\,Hz and between 40\,Hz and 6\,kHz for KAGRA and GEO, respectively. 
In addition, a cleaning process \cite{PhysRevD.101.102006} using auxiliary channels is applied to the GEO data to remove some bilinear noise from the gravitational wave strain data.

We have observed many short, transient noise fluctuations, known as glitches, in each detector.
During the joint observing period, the median rates of glitch triggers generated by the data-monitoring 
program  \omicron~\cite{Robinet:2015,Robinet:2020lbf} with \ac{SNR} larger than 6.5 were 10.3 per 
minute for GEO and 6.8 per minute for KAGRA. 
These values are significantly larger than the glitch rates during the first and the second parts of O3 (O3a and O3b) of LIGO--Virgo, which were 0.29--0.32 per minute, 1.1--1.2 per minute and 0.47--1.1 per minute for LIGO--Hanford, LIGO--Livingston and Virgo, respectively~\cite{LIGOScientific:2020ibl,LIGOScientific:2021djp}.
On the other hand, the glitch rates of GEO and KAGRA were comparable with the rate of 14 per minute in Virgo during the second observing run (O2)~\cite{LIGOScientific:2020ibl}, which was the first observing period for the Advanced Virgo project.
The investigation of sources of glitches in GEO and KAGRA is ongoing by identifying statistical coincidences 
and physical couplings  between the auxiliary channels and the strain channel.

One method to reduce the impact of glitches on \ac{GW} searches is through the use of \textit{data-quality flags}, 
lists of time segments that identify the status of detectors or the likely presence of a particular instrumental artefact. 
Three categories of data-quality flags are used in GW searches~\cite{LIGOScientific:2016gtq,LIGO:2021ppb}. 
Category 1 flags indicate that the data have been severely impacted by noise and should not be used 
for astrophysical searches. 
Category 2 flags indicate that the data are predicted to contain non-Gaussian artefacts based on 
glitches in auxiliary channels and known physical couplings to the strain data. 
Category 3 flags indicate that the data are predicted to contain non-Gaussian artefacts based on 
glitches in auxiliary channels and statistically significant correlations between glitches in auxiliary 
channels and glitches in the strain data. 
GEO has introduced data-quality flags corresponding to Category 1 and Category 3. 
KAGRA had not introduced data-quality flags by the time of the joint run; 
they are planned to be introduced before the next observing run. 

\section{All-sky binary search} 
\label{sec:gstlal}

To search for \ac{CBC} signals,
we first perform a matched-filter search
and then rank candidate events with a multi-dimensional classifier
using the \gstlal library~{\cite{Messick:2016aqy, Sachdev:2019vvd, Hanna:2019ezx}.
Because of their short duration, high-mass binary coalescences
are difficult to distinguish from glitches.
In this search, it was found that the brief observation period did
not provide a sufficiently large data set to train the ranking statistic,
leading to noise features being incorrectly assigned high statistical significance.
In contrast, because of their longer duration, \ac{BNS} waveforms
are easier to distinguish from noise transients,
and despite the short observation period there is sufficient data
to train the \gstlal detection system to perform well for this class of GW source.
For this reason we restrict the search to \ac{BNS} sources only.

Except for restricting the mass parameter range to \ac{BNS} sources, the \gstlal
configuration for this search is
the same as those for our most recent GW transient catalogs,
GWTC-2.1~\cite{LIGOScientific:2021usb} and GWTC-3~\cite{LIGOScientific:2021djp},
and for the O3a subsolar-mass binary search~\cite{LIGOScientific:2021job}, with one change:
the event clustering based on the matched-filter \ac{SNR}
is disabled, and instead a data reduction step based on \ac{SNR}
and the signal-consistency test statistic is newly introduced.
This change improves the GEO--KAGRA sensitive range by approximately $10\%$. 
This new finding will also help improve future LIGO--Virgo--KAGRA analyses. 

Matched filtering is done by comparing the data to a set of template waveforms called a template bank~\cite{Sathyaprakash:1991mt, Dhurandhar:1992mw, Owen:1995tm, Owen:1998dk}.
We use the same template bank
as was used in the first Advanced LIGO observing run (O1)~\cite{LIGOScientific:2016vbw, LIGOScientific:2016hpm}
but with the component masses restricted to the range 1\,\Msun to 3\,\Msun,
which conservatively covers the range expected for \acp{NS}~\cite{LIGOScientific:2021djp}.
Templates are parametrized in terms of their chirp mass ${\cal M}$ which is related to the individual component masses 
$m_1$, $m_2$ by ${\cal M} = (m_1 m_2)^{3/5} / (m_1 + m_2)^{1/5}$.
For templates with a chirp mass less than 1.73\,\Msun
the \texttt{TaylorF2} waveform approximant~\cite{Blanchet:2013haa, Dhurandhar:1992mw, Buonanno:2009zt} is used,
while for higher masses 
the reduced-order model of the \texttt{SEOBNRv4} approximant~\cite{Bohe:2016gbl} is used. 
This subset of the O1 template bank was tested against a set of \ac{BNS} signals with 
masses distributed uniformly across the search mass range and using noise power spectral 
densities typical of GEO and KAGRA during their joint run.
The fitting factor \cite{PhysRevD.52.605} was above 0.9 for $>$99\% of 
simulated signals, with the exceptions being simulations for the chirp mass larger than 2.4$M_\odot$. 
The fitting factor was above 0.97 (the threshold commonly used in LIGO--Virgo 
searches \cite{LIGOScientific:2021djp}) for all signals with chirp 
masses 
below 2\,\Msun, corresponding to component masses below 2.3\,\Msun 
for an equal-mass binary.

\gstlal defines triggers as the maximum of \ac{SNR}
over 1\,s windows which exceed a threshold of 4. 
It defines coincident triggers as triggers from each detector associated with the same template 
and with coalescence times within 32.5\,ms of each other. 
This time window accounts for the maximum light-travel time (27.5\,ms) between GEO and KAGRA 
as well as the uncertainty in the inferred coalescence time at each detector. 
Candidate events comprise both coincident triggers and non-coincident triggers. 
We define the network \ac{SNR} as the root-sum-square of the \ac{SNR}s for coincident triggers,
and simply the \ac{SNR} for non-coincident triggers.
We discard candidate events that have network \ac{SNR} 
below 7 because there are so many noise background events at those low \acp{SNR} 
that it is hard to distinguish true signals from noise.

\gstlal ranks candidate events based on the logarithm of
the likelihood ratio $\mathcal{L}$, which is a measure of how signal-like a given event is.
The likelihoods used in this analysis are constructed using the \ac{SNR},
a signal-consistency test,
the differences in time and phase between the triggers from different detectors when the candidate event consists of coincident triggers,
the information of which set of detectors (\{GEO\}, \{KAGRA\}, or \{GEO, KAGRA\}) form the event,
the sensitivity of the detectors to the exact template masses at the time of the event,
the rate of triggers in each of the detectors at the time of the event, and
the relative frequency with which signals are expected to be recovered by each template
given the assumption that astrophysical sources are distributed uniformly in the logarithm of the masses.

\gstlal uses Monte Carlo techniques to estimate
the distribution function $f(\ln \mathcal{L})$ for the log-likelihood ratios
assigned to candidates resulting from the noise process.
From $f(\ln \mathcal{L})$, the total number of candidates collected in the experiment,
and the experiment's duration we compute the mapping from
a log likelihood-ratio threshold $\ln \mathcal{L}_{\mathrm{th}}$
to the false-alarm rate, $\mathrm{FAR}(\ln \mathcal{L}_{\mathrm{th}})$,
which is the rate at which the noise process yields candidates at or above the given threshold.
  
\subsection{Search results}
\label{sec:gstlal_results}

For the GEO--KAGRA search, the total amount of data analyzed for each detector combination 
was 4.59 days for GEO-only, 0.90 days for KAGRA-only, and 6.21 days for 
two-interferometer observations, for a total of 11.70 days (0.032\,years).

Figure \ref{fig:gstlal_rate_vs_threshold} shows the event count
as a function of the threshold on the \ac{iFAR}. 
We see no significant deviation of the observed distribution from our noise model
and conclude that no signal of interest has been detected.
The most significant candidate is found as a coincident trigger in GEO and KAGRA
at April 20 2020 14:03:28 UTC 
with an \ac{iFAR} of 0.033\,years.

\begin{figure}[hbt]
    \centering
    \includegraphics[width=0.7\textwidth]{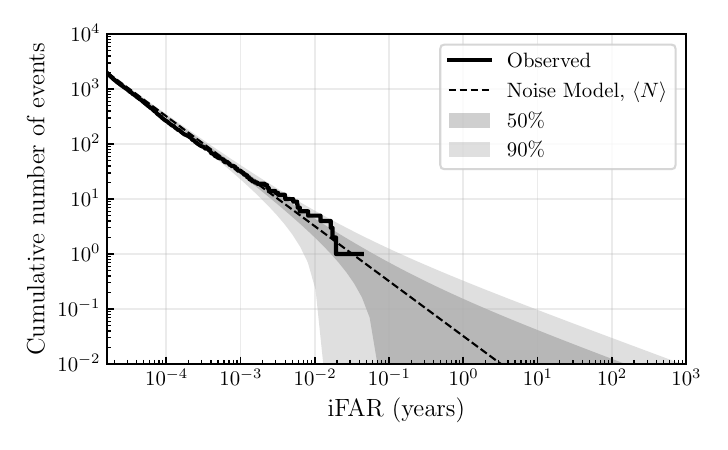}
    \caption{Event count versus threshold on iFAR.
      The predicted distribution due to noise is shown as the dashed line along with 
      its 50\% and 90\% statistical error regions.
      The observed distribution is shown as the solid line.}
    \label{fig:gstlal_rate_vs_threshold}
\end{figure}

\subsection{Search sensitivity}
\label{sec:gstlal_sensitivity}
We estimate the sensitive spacetime volume 
(product of sensitive volume and livetime) 
of this search to CBCs
by adding simulated signals to the data and repeating the analysis~\cite{Tiwari:2017ndi}. 
Since GEO and KAGRA were not sensitive enough to constrain the \ac{BNS} merger 
rate beyond the limits already set by LIGO and Virgo \cite{LIGOScientific:2021psn}, 
we do not use an astrophysically motivated distribution for \ac{BNS} masses. 
Instead, we measure the search sensitivity around a canonical \ac{BNS} mass of $1.4\,\Msun$. 
Specifically, the simulated signals are generated so that each component mass is normally distributed 
with a mean of $1.4\,\Msun$ and a standard deviation of $0.01\,\Msun$; 
\textit{i.e.}, according to $\mathcal{N}(1.4\,\Msun, [0.01\,\Msun]^2)$.
The waveform approximant used for the simulated signals is
\texttt{TaylorT4} to 3.5 post-Newtonian order~\cite{Buonanno:2002fy, Boyle:2007ft, Buonanno:2006ui, Blanchet:2001ax}.
The signals are spaced uniformly in time with an average spacing of 10\,s.
Their sources are distributed uniformly in distance between 0.1\,Mpc and 3\,Mpc 
and isotropically across the sky and in orientation.
Figure~\ref{fig:gstlal_vt} shows the sensitive spacetime volume 
as a function of the iFAR threshold.
This volume is computed by integrating detection efficiency over distance with appropriate weighting, 
where the efficiency is defined as the fraction of simulated signals
that exceed the iFAR threshold within each distance bin.
When we compute this fraction, we include GEO--only, KAGRA-only, and GEO--KAGRA times.
The spacetime volume is a decreasing function of the threshold, 
approximately $3 \times 10^{-2}$\,Mpc$^{3}$\,years to $2 \times 10^{-2}$\,Mpc$^{3}$\,years 
for \ac{iFAR}s from one per year to one per million years.
Figure~\ref{fig:gstlal_vt} also shows the equivalent sensitive range, defined 
as the radius of a sphere of the same average spatial volume, 
which may be compared to Figure~\ref{fig:spectra}. 
The \ac{iFAR} of the most significant candidate corresponds to a range of $\sim$0.6\,Mpc,
which can be taken as the approximate sensitive range of this analysis.

\begin{figure}[hbt]
    \centering
    \includegraphics[width=0.7\textwidth]{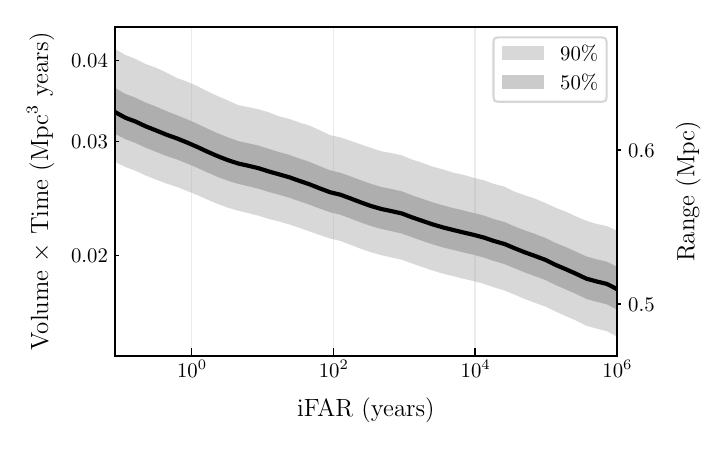}
    \caption{Sensitive spacetime volume to \ac{BNS} coalescences 
      with component masses drawn from $\mathcal{N}(1.4\,\Msun, [0.01\,\Msun]^2)$ 
      as a function of the threshold on \ac{iFAR} for the {\gstlal} binary search.
      The equivalent range (right axis) is also shown. 
      The bands show the 50\% and 90 \% error regions,
      estimated as the Wilson score interval~\cite{Wilson:1927}.}
    \label{fig:gstlal_vt}
\end{figure}
\section{All-sky burst search}
\label{sec:cwb}

The search pipeline \ac{CWB}~\cite{klimenko2016method,drago2021coherent} is an algorithm for the detection 
and reconstruction of \ac{GW} transient signals with durations of typically up to a few seconds.
The algorithm searches for coincident excess signal power in a network of \ac{GW} detectors 
without assuming specific waveform models, and therefore is suitable for searching for \ac{GW} 
transients from a range of different sources. 
It is used in all-sky burst searches 
\cite{LIGOScientific:2012cxo, LIGOScientific:2016kum, LIGOScientific:2019ppi}, 
as well as for example in searches for \ac{GW}s from binary coalescences 
\cite{LIGOScientific:2018mvr,LIGOScientific:2020ibl, LIGOScientific:2021djp} 
and core-collapse supernovae~\cite{LIGOScientific:2019ryq}.

Analyses with \ac{CWB} are performed in a wavelet domain~\cite{Necula:2012zz} on normalized data transformed at various resolution levels. 
Wavelets with amplitudes above the typical fluctuations of detector noise are selected and grouped into clusters. 
Clusters that are correlated in multiple detectors are identified as coherent events. 
For coherent events, waveforms are reconstructed based on maximum-likelihood-ratio statistics~\cite{klimenko2016method}. 
Events are ranked by their coherent network SNR~$\eta_\mathrm{c}$~\cite{klimenko2016method} and those with $\eta_\mathrm{c}>5$ are stored for further processing. 

Due to the high rates of glitches and a large number of noise coincidences found 
in the GEO--KAGRA network, we apply an additional constraint in which only one polarization component 
of a \ac{GW} candidate event is reconstructed. 
This constraint has been employed in other LIGO and Virgo searches~\cite{LIGOScientific:2012cxo, LIGOScientific:2016kum, LIGOScientific:2019ppi, LIGOScientific:2019ryq}. 
It is effective in mitigating the background event rate, and allows the analysis to search for the 
\ac{GW} polarization to which the network has maximum sensitivity from each sky direction~\cite{Klimenko:2005xv,Sutton:2009gi}. 
However, for non-aligned detectors, such as GEO and KAGRA, 
each detector can be sensitive to different polarizations at any given sky location. 
In this case the constraint may lead to the rejection of real events. 
Also, where the network is sensitive to both polarizations, a significant portion of the signal energy contributes to the noise estimate. 
In extreme cases, the contribution may be so large that the signal becomes undetectable. While reconstructing both polarizations may help reduce false negative rate, 
lifting this constraint would increase substantially the background event rate as well as the computational cost.

To reduce further the rate of noise events falsely identified as \ac{GW} signals, we apply additional selection cuts. 
In this work we use the network correlation coefficient $c_{\rm c}$~\cite{klimenko2016method}, 
which is a ratio between correlated and total energy of the signal. 
\ac{GW} signals have $c_{\rm c}\approx 1$; we exclude events with $c_{\rm c} < 0.55$.
We also employ the effective number of time--frequency resolution levels used for event detection and 
waveform reconstruction~\cite{Mishra:2021tmu} $n_f$.  In total 14 resolution levels are used in this analysis.
For noise events the typical values of $n_f$ are low; we exclude events with $n_f < 8.9$.
These thresholds are selected based on separating background events and simulated signals 
(described in Sections~\ref{sec:bkg} and~\ref{sec:searchsen}).
We further exclude events with central frequency in the range 118--124\,Hz 
because a significant number of background events with central frequency near 120\,Hz were observed during the run.
An analysis of these glitches with \omicron (Section~\ref{sec:run}) 
indicates they are likely associated with a single unknown noise source in KAGRA.

\subsection{Background and search results}
\label{sec:bkg}

Given that the GEO--KAGRA network sensitivity is limited both at frequencies $\lesssim100$\,Hz and $\gtrsim1$\,kHz (see Figure~\ref{fig:spectra}), 
our analysis spans the frequency range of 64--1024\,Hz.
The data is down-sampled and periods of poor data quality are removed, similar to the all-sky searches for burst signals in O1 
and O2 ~\cite{LIGOScientific:2016kum, LIGOScientific:2019ppi}. 
Intervals with at least $600$\,s of continuous coincident data are required, and the total analysed coincident time between GEO and KAGRA is equal to $4.38$ days.
The background event distribution is estimated by artificially time--shifting the data from one detector with respect to the other. 
The time shifts are multiples of 1\,s, larger than the time required for a \ac{GW} signal to travel between 
the detectors so that any identified signal is not of astrophysical origin.
In total, a background livetime of $7.2$ years is obtained.

Figure~\ref{fig:far} shows the background distribution before and after application of 
the $c_\mathrm{c}$, $n_f$ and central-frequency selection cuts.
The post-selection-cut distribution is considered the background distribution of events for this analysis.

\begin{figure}[hbt]
    \centering
    \includegraphics[width=0.7\textwidth]{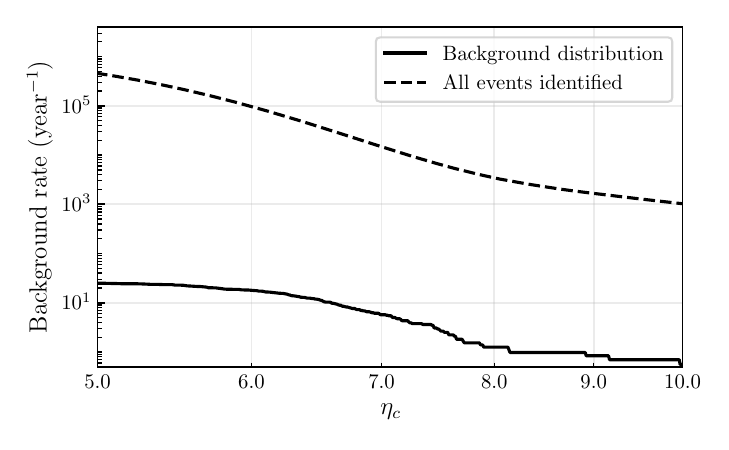}
    \caption{The rate of background events as a function of coherent network SNR $\eta_\mathrm{c}$ for the \ac{CWB} all-sky burst search. 
    The dashed line shows the rates for all the events.
    The solid line shows the rate after application of the $c_\mathrm{c}$, $n_f$ and central-frequency selection cuts.
    }
    \label{fig:far}
\end{figure}

Figure \ref{fig:cumulativeevents} shows the event count
as a function of the threshold on the \ac{iFAR}. 
Only one candidate event is identified, at April 12 2020 18:10:15 UTC with an \ac{iFAR} of $0.097$ years. 
It is consistent with the background and is not significant enough to be considered a \ac{GW} event.

\begin{figure}
\centering
\includegraphics[width=0.7\textwidth]{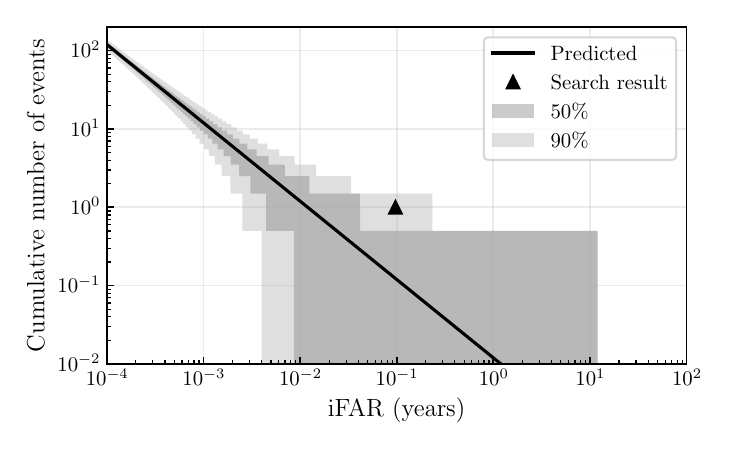}
\caption{Cumulative number of events with central frequency in 64--1024\,Hz versus \ac{iFAR} found by the \ac{CWB} all-sky burst search. 
Only a single event is identified (triangle).
The shaded regions show the $50\%$ and $90\%$ Poisson uncertainties. 
\label{fig:cumulativeevents}}
\end{figure}

\subsection{Search sensitivity}
\label{sec:searchsen}
We estimate the search sensitivity to potential \ac{GW} transients by 
adding simulated signals to the detector data and repeating the analysis. 
Similarly to other observing runs~\cite{LIGOScientific:2016kum, LIGOScientific:2019ppi, LIGOScientific:2012cxo}, 
we use a variety of \textit{ad hoc} waveforms including \ac{SG}, \ac{GA}, and band limited \ac{WNB}, with 
frequencies and duration spanning a range of possible values. 
\ac{SG} signals are defined by their central frequency $f_0$ and quality factor $Q$, which determines the duration of the signals. 
The \ac{GA} signals are described by their duration $\tau$. The \ac{WNB} signals are described by their lower frequency bound $f_{\mathrm{low}}$, 
bandwidth $\Delta f$, and duration $\tau$. 
The parameter values chosen are listed in Table~\ref{table:simulated_waveforms}.
In addition to these \textit{ad hoc} signals, two astrophysically motivated signals are used: 
the reconstructed signal of \ac{GW}150914~\cite{GW150914} and a simulated 
core-collapse supernova waveform referred to as SFHx~\cite{kuroda2016new}.

The simulated signals are distributed uniformly over the sky and in polarization angle.
For \ac{SG} waveforms, we use both elliptical and circular polarizations: 
the sources of circular \ac{SG}s are assumed to be optimally oriented 
while the sources of elliptical \ac{SG}s have isotropically distributed orientations.
\ac{GA} waveforms are linearly polarized, 
while \ac{WNB} waveforms have uncorrelated equal-amplitude polarizations.
For SFHx, we use the optimal orientation as the waveform is only available at this observing angle.
Each signal is simulated at a wide range of amplitudes, characterized by the root-sum-squared strain $h_{\mathrm{rss}}$:
\begin{equation}\label{eq:hrss}
 h_{\mathrm{rss}} = {\bigg\{\int_{-\infty}^{\infty} [h_+^2(t)+h^2_\times(t)] \, \mathrm{d}t\bigg\}}^{1/2}.
\end{equation}
These signals are then recovered using the search method described above and 
the detection efficiency is defined as the fraction of signals that produce 
an event which passes the selection cuts and has an \ac{iFAR} $\ge1$\,year.

Table~\ref{table:simulated_waveforms} shows for each waveform type the $h_{\mathrm{rss}}$ amplitude at which the detection efficiency reaches 50\% and
90\%. As mentioned earlier, the constraint employed in \ac{CWB} affects the sensitivity of networks of two detectors.
This effect is more prominent when the reconstructed waveform energy is distributed across different polarization components.
As a result, the detection efficiencies for these waveforms are less than $90\%$ even for large values of $h_{\mathrm{rss}}$. 
For these waveforms, we put N/A in the column corresponding to $90\%$ detection efficiency.
These $h_{\mathrm{rss}}$ limits follow the network noise spectra (Figure~\ref{fig:spectra}).
\begin{table}[]
\centering
\begin{threeparttable}
\begin{tabular}{ccc}
\hline
\hline
\textbf{Morphology}                                                                &  50\%               &     90\%                \\
\hline
\multicolumn{1}{c}{\textbf{Gaussian pulses (linear)}}                              &   \multicolumn{2}{l}{$h_\mathrm{rss}$ ($10^{-20}\,\text{Hz}^{-1/2 }$)}     \\
$\tau = 0.1$\,ms                                                                   &  $5.3$              &     N/A                 \\                                    
$\tau = 2.5$\,ms                                                                   &  $15.0$             &     N/A                 \\
\multicolumn{1}{c}{\textbf{sine--Gaussian wavelets (circular)}}                                                                    \\
$f_0 = 100$\,Hz, $Q = 9$                                                           &  $4.9$              &     $11.0$                 \\
$f_0 = 235$\,Hz, $Q = 9$                                                           &  $1.0$              &     $1.9$                 \\
$f_0 = 361$\,Hz, $Q = 9$                                                           &  $0.9$              &     $1.7$                 \\
\multicolumn{1}{c}{\textbf{sine--Gaussian wavelets (elliptical)}}                                                                  \\
$f_0 = 70$\,Hz, $Q = 3$                                                            &  $28.0$             &     $94.0$              \\
$f_0 = 153$\,Hz, $Q = 9.0$                                                         &  $4.0$              &     $14.0$               \\
$f_0 = 235$\,Hz, $Q = 100$                                                         &  $1.4$              &     $4.7$               \\
$f_0 = 554$\,Hz, $Q = 9.0$                                                         &  $1.5$              &     $4.3$               \\
$f_0 = 849$\,Hz, $Q = 3$                                                           &  $3.5$              &     $12.0$               \\
\multicolumn{1}{c}{\textbf{White--Noise Bursts}}                                                                                   \\
$f_{\text{low}}$ = 150\,Hz, $\Delta f = 100$\,Hz, $\tau = 0.1$\,s                  &  $1.9$              &     N/A                 \\
$f_{\text{low}}$ = 300\,Hz, $\Delta f = 100$\,Hz, $\tau = 0.1$\,s                  &  $1.1$              &     N/A                 \\
$f_{\text{low}}$ = 700\,Hz, $\Delta f = 100$\,Hz, $\tau = 0.1$\,s                  &  $1.2$              &     N/A                 \\
\textbf{Astrophysical Signals}                                                                &   \multicolumn{2}{l}{distance (kpc)}                          \\
\ac{GW}150914                                                                      &  $809$  &     N/A                 \\
Supernova SFHx                                                                     &  0.08   &     $0.01$               \\
\hline
\hline
\end{tabular}
\caption{
The \ac{GW} morphologies used to quantify the search sensitivity.
The first column shows the waveforms used. 
The second and third columns show the $h_{\text{rss}}$ values at which 
$50\%$ and $90\%$ detection efficiencies are achieved at an \ac{iFAR} of  1\,year. 
For the astrophysical waveforms the second and third columns show the luminosity distance at which 
these efficiencies are achieved.
}
\label{table:simulated_waveforms}
\end{threeparttable}
\end{table}

Assuming isotropic and narrow-band emission by a source, the energy emitted in \ac{GW}s is given by~\cite{LIGOScientific:2012cxo}:
\begin{equation}\label{eq:eng}
E_{\mathrm{GW}} = \frac{\pi^2 c^3 }{G} \, r^{2}f^{2}_{0}h_{\text{rss}}^2 \, ,
\end{equation}
where $r$ is the distance to the source and $f_{0}$ is the central frequency.
This equation is valid for unpolarized signals such as \ac{WNB}s, while for \ac{SG} signals, 
the rotating system emission has to be accounted for by multiplying the right-hand side of Eq.~\eqref{eq:eng} by a factor of 2/5~\cite{Sutton:2013ooa}.
Using Eq.~\eqref{eq:eng} and the $h_{\text{rss}}$ limits from Table~\ref{table:simulated_waveforms}, 
we can estimate the minimum energy needed to be radiated by a population of standard-candle
sources at a distance of $r = 10$\,kpc to give a $50\%$ detection efficiency.
The results are shown in Figure~\ref{fig:sen}.
Again, the general behavior is determined by the power spectral density of the network (Figure~\ref{fig:spectra}).

\begin{figure}
\centering
\includegraphics[width=0.7\textwidth]{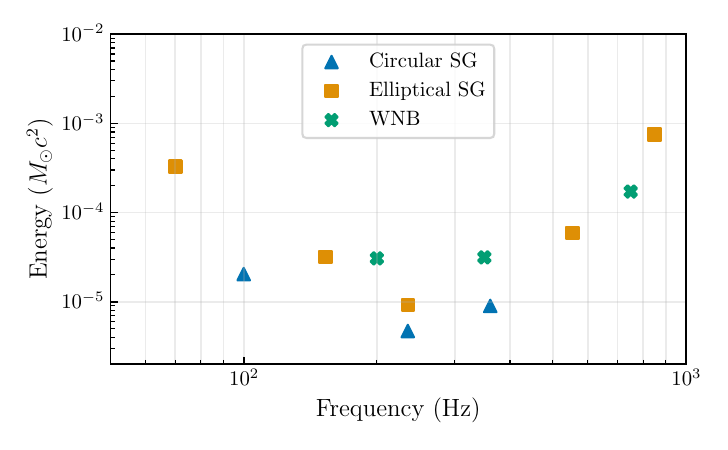}
\caption{The \ac{GW} emitted energy in units of solar masses ($M_\odot c^2$)
that correspond to a $50\%$ detection efficiency with \ac{CWB} at an \ac{iFAR} of $\geq 1$ year, for a source located at $10$ kpc. 
The circular \ac{SG} waveforms are indicated by triangles, the elliptical  \ac{SG} waveforms by squares, 
and the \ac{WNB} waveforms by crosses.
\label{fig:sen}}
\end{figure}

\section{Gamma-ray burst analyses}
\label{sec:grb}

\acp{GRB} are targets of interest in \ac{GW} astronomy because the astrophysical processes that power them, specifically massive stellar core collapse~\cite{2017hsn..book.1671K,Ott_2009,Fryer:2001zw,vanPutten:2003hd,Piro:2006ja,Corsi:2009jt} and \acp{CBC}~\cite{LIGOScientific:2017zic}, may also emit detectable \acp{GW}.
By targeting \acp{GRB} with tailored search methods we can potentially detect weaker associated \acp{GW} than would be identified with non-targeted analyses \cite{Sutton:2009gi,Williamson:2014wma}.

\acp{GRB} display a bimodality in their joint duration-spectral-hardness distribution~\cite{Kouveliotou:1993yx}.
Long-soft \acp{GRB} (duration $\gtrsim$ 2\,s) are associated with massive stellar core 
collapse~\cite{Galama:1998ea,Hjorth:2003jt,Stanek:2003tw}.
The physics governing the bulk motion of matter during these events is complex,
so we do not have robust models of the resulting \ac{GW} emission, though a number of
speculative models for strong \acp{GW} emission have been proposed, such as long-lived bar-mode instabilities 
and disk fragmentation instabilities~\citep{Fryer:2001zw,vanPutten:2003hd,Piro:2006ja,Corsi:2009jt}. 
We therefore use a minimally modeled search algorithm \Xpipeline~\cite{Sutton:2009gi,Was:2012zq} to target these \acp{GRB}.

Short-hard \acp{GRB} (duration $<$ 2\,s) can be produced by \ac{NS} binary coalescences, a connection that 
was long proposed~\cite{Blinnikov:2018boq,Eichler:1989ve,Paczynski:1991aq,Narayan:1992iy} and
observationally confirmed by the multimessenger studies of GW170817/\mbox{GRB\,170817A}~\cite{TheLIGOScientific:2017qsa,LIGOScientific:2017zic,GBM:2017lvd,Lazzati:2017zsj,Alexander:2018dcl,Mooley:2018dlz,Ghirlanda:2018uyx,Fong:2019vgn}.
We therefore target them with a modeled \ac{CBC} search algorithm \pygrb~\cite{Harry:2010fr,Williamson:2014wma}
in addition to the more generic minimally modeled \Xpipeline.

During the joint GEO--KAGRA run, \nCoincGRB \acp{GRB} were detected coinciding with science data taking in both GEO and KAGRA; see Table~\ref{tab:grbs}.
Our minimally modeled search algorithm was able to analyze all of these given its data requirements.
GRB\,200415A and GRB\,200420A were short duration, therefore our modeled search algorithm was also used to analyze them.
GRB\,200415A was subsequently associated \cite{Roberts:2021udn,Svinkin:2021wcp} with a magnetar giant flare in the nearby galaxy NGC253 at 3.5\,Mpc based on its sky position, temporal and spectral properties and inferred energy.
All \ac{GRB} properties were taken from the \textit{Fermi} \ac{GBM} Catalog~\cite{fermigbmcatalog,vonKienlin:2020xvz,Bhat:2016odd,vonKienlin:2014nza,Gruber:2014iza}, with one exception: the minimally modeled analysis of GRB\,200415A took sky position data from a preliminary IPN triangulation~\cite{gcn27595} for practical reasons.
Given the coarse angular sensitivity of the \ac{GW} detector network, the very small difference does not affect the results in any significant way.

\begin{table}[]
    \centering
    \begin{threeparttable}
        \begin{tabular}{l l l l}
            \hline
            \hline
            GRB Name & Data Source & Type & Analysis \\
            \hline
            200412A & \textit{Fermi}-GBM & Long & \Xpipeline \\ 
            200415A & \textit{Fermi}-GBM, IPN & Short & \Xpipeline \& \pygrb \\  
            200418A & \textit{Fermi}-GBM & Long & \Xpipeline\\     
            200420A & \textit{Fermi}-GBM & Short & \Xpipeline \& \pygrb\\ 
            \hline
            \hline
        \end{tabular}
        \caption{
            \acp{GRB} observed during GEO--KAGRA run times when both detectors were taking science-quality data. 
            GRB\,200415A and GRB\,200420A were short-duration \acp{GRB}, and so are analysed by both searches. 
        }
        \label{tab:grbs}
    \end{threeparttable}
\end{table}

\subsection{Binary coalescence search targeting short GRBs}
\label{sec:pygrb}
By targeting the times and sky positions of short \acp{GRB}, we can perform a deep, coherent
matched filter analysis for associated \acp{GW} from \ac{BNS} and \ac{NSBH} binaries.
This analysis is called \pygrb~\cite{Harry:2010fr,Williamson:2014wma}, and forms part of the larger
\pycbc analysis toolkit~\cite{pycbc} with key components in the \LALSuite library~\cite{lalsuite}.
This approach has been used in many previous observing runs of the LIGO and Virgo detectors~\cite{LIGOScientific:2016akj,LIGOScientific:2019obb,LIGOScientific:2020lst,LIGOScientific:2021iyk} and here we deploy a \pygrb analysis
that is functionally identical to that used in the most recent LIGO--Virgo analyses~\cite{LIGOScientific:2020lst,LIGOScientific:2021iyk},
with only some changes to the configuration that are appropriate for the data being analyzed, as outlined below.

The \pygrb search performs a matched filter coherently across the operational \ac{GW} detector network around the time of each short \ac{GRB}.
In this analysis we filter in the frequency range 40--1000\,Hz with a bank of template waveforms~\cite{Owen:1998dk,Capano:2016dsf} generated with an aligned-spin point-particle model, \texttt{IMRPhenomD}, that includes inspiral, merger, and ringdown phases~\cite{Husa:2015iqa,Khan:2015jqa}.
The bank includes waveforms representing \ac{BNS} and \ac{NSBH} systems, where \acp{NS} have 
dimensionless spins \review{$\leq 0.05$}.
\footnote{The fastest known spinning pulsar has a dimensionless spin magnitude of $\sim$0.4~\cite{Hessels:2006ze}
and masses bound by $[1.00, 2.83] \, \Msun$ and \acp{BH} have dimensionless spins \review{$\leq 0.998$} and masses bound by $[2.83, 25.00] \, \Msun$.
We restrict our template bank to \ac{NS} spin magnitudes of $\le0.05$ because it has been demonstrated \cite{Nitz2015,LIGOScientific:2016hpm} that due to the balance between signal recovery and false alarm rate, the overall search sensitivity for \ac{BNS} systems with spins $<$0.4 is larger when the template bank is restricted to spins $<$0.05 than when it is expanded to include spins $<$0.4.}
Within these bounds, \ac{NSBH} templates are further constrained to the region in parameter space where the combination of masses and spins could give rise to tidal disruption of the \ac{NS}, and therefore potentially produce a \ac{GRB} following \cite{Foucart:2012nc,Pannarale:2014rea}, assuming a very stiff 2H equation of state \cite{Kyutoku:2010zd} and requiring a non-zero remnant mass.
Additionally, we place an inclination constraint motivated by the expected \cite{SGRBsurvey2012,Fong:2013lba,Panaitescu:2005er,Troja:2016vzw} 
small inclination angles for \ac{GRB} progenitors due to \ac{GRB} beaming.
This is imposed by filtering with only circularly polarized templates~\cite{Williamson:2014wma}, corresponding to binary systems with inclination angles $\theta_{JN}$ between the total angular momentum axes $\hat{J}$ and the line-of-sight $\hat{N}$ of $0\,\deg$ or $180\,\deg$.
This constraint improves sensitivity to signals with small inclinations ($\lesssim 30\,\deg$ or $\gtrsim 150\,\deg$).

We tile the reported sky error region of each \ac{GRB} and filter at each sky point with our constrained template bank~\cite{Williamson:2014wma} to obtain a coherent \ac{SNR} statistic for the network.
We place thresholds of 4 on single detector \acp{SNR} and 6 on the coherent network \ac{SNR}.
Surviving triggers are then re-weighted or cut according to signal consistency checks~\cite{Allen:2005fk,Harry:2010fr,Williamson:2014wma}, to produce the search detection statistic.

We consider a 6\,s window spanning $[-5,+1) \, \mathrm{s}$ about the reported \ac{GRB} Earth-crossing time as the on-source where an associated \ac{GW} event may be found.
This is compared to an off-source window that is used to characterize the search background,
which typically contains up to $\sim 90\,\mathrm{min}$ of data surrounding the on-source time.
The loudest (most significant) candidate event in the on-source, as defined by the detection statistic, is compared to a list containing the most significant background events from each of the 6\,s background trials within the off-source.
Additional background trials are obtained by time shifting the data streams relative to one another by amounts greater than the light travel time between the detectors~\cite{Williamson:2014wma}, similar to the approach described in Section~\ref{sec:cwb}.
This comparison between on-source and background trials results in a $p$-value for the candidate on-source event.

The short \ac{GRB} triggers during the analysis period with available data from both interferometers were GRB\,200415A and GRB\,200420A.
The loudest candidates within the on-source windows had $p$-values of \pvalPYGRBAprFifteen and \pvalPYGRBAprTwenty respectively, 
consistent with being due to background noise.

The sensitivity of the search is evaluated through the use of simulated \ac{GW} signals inserted throughout the off-source data and spread across the region(s) of the sky corresponding to the positional uncertainty of the \ac{GRB} trigger.
These simulated signals correspond to events drawn from three potential astrophysical populations:  \ac{NSBH} with aligned spins, \ac{NSBH} with isotropically oriented spins, and \ac{BNS} with isotropically oriented spins.
We draw \ac{NS} masses from normal distributions centered on 1.4 \Msun with standard deviations of $0.2 \Msun$ and $0.4 \Msun$ for \ac{BNS} and \ac{NSBH} systems respectively~\cite{Ozel:2012ax,Kiziltan:2013oja}, limited within the range $[1.0, 3.0] \Msun$.
The wider \ac{NSBH} distribution reflects the greater uncertainty surrounding \ac{NSBH} system properties.
\ac{NS} dimensionless spin magnitudes are drawn uniformly in the range $[0, 0.4]$, with the upper limit corresponding to the fastest spinning pulsar observed~\cite{Hessels:2006ze}.
\ac{BH} masses are drawn from ${\cal N}(10 \Msun, [6 \Msun]^2)$ limited within the range $[3, 15] \Msun$ and dimensionless spin magnitudes uniformly in the range $[0, 0.98]$~\cite{Miller:2014aaa}.
Spins are isotropically oriented except for the aligned spin \ac{NSBH} population.
Inclination angles $\theta_{JN}$ are drawn uniformly in $\cos \theta_{JN}$ for $\theta_{JN} \in [0, 30^{\circ}] \cup [150^{\circ}, 180^{\circ}]$.
\ac{NSBH} systems are then rejected if they do not meet the same \ac{NS} disruption condition as applied to the template bank~\cite{Foucart:2012nc,Pannarale:2014rea}.
\ac{NSBH} signals are generated with a point-particle effective-one-body model for the inspiral--merger--ringdown phases that incorporates orbital precession effects and is tuned to numerical-relativity simulations, \texttt{SEOBNRv3}~\cite{Pan:2013rra,Taracchini:2013rva,Babak:2016tgq}.
\ac{BNS} signals are generated with a time-domain approximation to 3.5 post-Newtonian order for the inspiral phase, \texttt{SpinTaylorT2}~\cite{Sathyaprakash:1991mt,Blanchet:1996pi, Mikoczi:2005dn,Arun:2008kb,Bohe:2013cla,Bohe:2015ana,Mishra:2016whh}.

In the case of no compelling candidate event being identified in the on-source, these simulated signals allow for exclusion distances to be quoted.
A 90\% exclusion distance corresponds to the distance within which 90\% of a population of simulated signals were recovered with a detection statistic at least as large as the loudest on-source candidate event; at greater distances the recovered fraction of signals drops.
For GRB\,200415A we report 90\% exclusion distances of \DBNSAprFifteen for \ac{BNS} systems, \DNSBHGenAprFifteen for isotropically spinning \ac{NSBH}, and \DNSBHAliAprFifteen for aligned-spin \ac{NSBH}.
At a distance of 3.5\,Mpc, corresponding to NGC~253, exclusion confidences for these three populations are \pcBNSAprFifteenNGC, \pcNSBHGenAprFifteenNGC, and \pcNSBHAliAprFifteenNGC respectively, too low to be able to confidently exclude any such binary merger as the progenitor of GRB\,200415A.
These exclusion curves are shown in Figure~\ref{fig:200415367_excl}.
For GRB\,200420A we report 90\% exclusion distances of \DBNSAprTwenty for \ac{BNS} systems, \DNSBHGenAprTwenty for isotropically  spinning \ac{NSBH}, and \DNSBHAliAprTwenty for aligned-spin \ac{NSBH}.
The injection recovery was limited in this case by the large sky error of the \ac{GRB}.
The reported \ac{GBM} $1\sigma$ statistical uncertainty (averaged over the error ellipse~\cite{vonKienlin:2014nza}) was $27.3\,\deg$~\cite{gcn27609}, and was used to generate a two-dimensional normal distribution on the sky from which injection sky positions were drawn.
This resulted in a population of injections spanning a large area on the sky within which the interferometer sensitivities varied significantly, including regions with severely reduced range.
As a result, a non-negligible fraction of nearby injections were undetectable.

\begin{figure}
    \centering
    \includegraphics[width=0.7\textwidth]{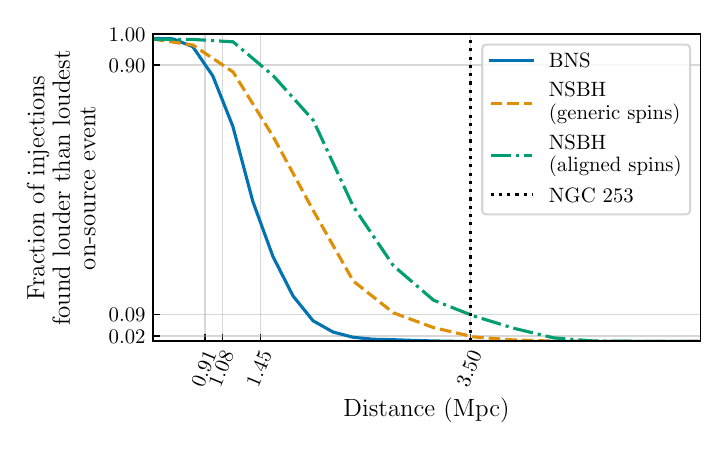}
    \caption{
        Exclusion distance curves for GRB\,200415A.
        We show the curves for each of our three injection populations: \acp{BNS} (blue solid), isotropically spinning \acp{NSBH} (orange dashed), and aligned-spin \acp{NSBH} (green dot-dashed).
        The respective 90\% confidence exclusion distances of \DBNSAprFifteen, \DNSBHGenAprFifteen, and \DNSBHAliAprFifteen are marked, as are the confidence levels corresponding to the distance to NGC~253 (3.5\,Mpc; black dotted), which are 0\%, 2\%, and 9\% respectively.
        Thus the search sensitivity is not sufficient to confidently exclude a binary merger in NGC~253 as the progenitor based on the available \ac{GW} data.
        \label{fig:200415367_excl}
    }
\end{figure}

\subsection{Search for generic bursts associated with GRBs}

\xpipeline  \cite{Sutton:2009gi,Was:2012zq} is an analysis package that combines data from multiple 
detectors coherently to detect minimally modeled \ac{GW} transient signals associated with events such as 
\ac{GRB}s, core-collapse supernovae, and fast radio bursts. It is used regularly for 
such searches of LIGO--Virgo data \cite{LIGOScientific:2016jvu,LIGOScientific:2016rmm,LIGOScientific:2019ccu,
LIGOScientific:2016akj,LIGOScientific:2017zic,LIGOScientific:2019obb,LIGOScientific:2020lst}.

For each \ac{GRB} \xpipeline constructs a grid of sky positions covering that \ac{GRB}'s sky localisation error box. 
For this analysis linear grids are used, which have been shown to be a computationally efficient way to 
cover large error boxes for two-detector networks without significant loss of sensitivity \cite{LIGOScientific:2014zwy}.
For each grid point a coherent analysis is performed.  The frequency range of the search is increased from the 
standard values of [20, 500]\,Hz to [30,1100]\,Hz to account for GEO's better sensitivity at higher frequencies. 
The on-source window is \mbox{[$-$600\,s, $+$max(60\,s,\textit{$T_{90}$})]} 
about the \ac{GRB} Earth-crossing time, where \textit{$T_{90}$} is the reported \ac{GRB} duration.
This window is large enough to account for any reasonable time delay between the \ac{GW} and gamma-ray 
emission \cite{koshut1995gamma, aloy2000relativistic, macfadyen2001supernovae, zhang2003relativistic, 
lazzati2005precursor, wang2007grb, burlon2008precursors, burlon2009time, lazzati2009very, vedrenne2009gamma}. 
An exception to this window choice is made for GRB\,200415A, for which KAGRA was not operating in a stable locked state 
until less than 600\,s before the \ac{GRB} event. For this \ac{GRB} we use an on-source window of \mbox{[$-$519, $+$60]\,s}. 
The off-source window consists of all data within $\pm$90\,min of the \ac{GRB}, including time shifts similar to those used by \ac{CWB}. 
The total amount of off-source data analysed is between $5\times10^3$ and $3\times10^4$ times 
the on-source duration for each \ac{GRB}, allowing $p$-values of order $10^{-4}$ to be measured.
Finally, simulated signals are added to the on-source window; these are used both for estimating 
the sensitivity of the search and for automated tuning of \xpipeline's background rejection tests. 

The same procedure is used for the on-source, off-source, and simulation analyses. 
The data are whitened, then Fourier transformed with transform durations of 
[1/256, 1/128, \ldots, 2]\,s. The Fourier-transformed data are combined to form time--frequency 
maps for each detector. From these maps the highest 1\% of pixels are grouped into clusters. 
For each cluster the data from the different detectors is combined in multiple combinations 
to estimate the signal energy consistent with different GW polarizations and to give 
various measures of correlation between detectors. 
When clusters from different sky positions or Fourier transform durations overlap in time-frequency, 
the most significant is retained.
The clusters are then checked for coherency between detectors to reduce the background. 
The thresholds for these background rejection tests are selected to maximise the detection efficiency at a
user-specified false-alarm probability ($10^{-4}$ for this analysis), using a subset of the off-source and simulation clusters.
The optimised thresholds are then applied to the on-source clusters and to the remaining off-source 
and simulation clusters. The surviving on-source clusters are our candidate events. Each is assigned a 
$p$-value by comparing to the distribution of surviving off-source events. 
The sensitivity as a function of signal amplitude or source distance is evaluated as the fraction of 
simulated signals that give surviving events with $p$-values lower than the lowest $p$-value of the on-source events.

Of the four \acp{GRB} analysed, the lowest $p$-value for any on-source event was $p=$~\pvalBurstLowest for 
\nameBurstLowest.  This is consistent with the null hypothesis given the number of \acp{GRB} analysed. 
We therefore conclude there is no evidence for GW emission associated with any of the four \acp{GRB} analysed. 
Figure~\ref{fig:xpipeline} shows the 90\% confidence level lower limit on the distance for each of the \acp{GRB} 
for several emission models: the accretion-disk instability model A of \cite{vanPutten:2001sw,vanPutten:2014kja}; 
and circularly polarized sine--Gaussian~\cite{LIGOScientific:2016akj} signals with central frequencies of 150\,Hz, 
500\,Hz, and 1000\,Hz where we assume an energy emission of $10^{-2}\,M_\odot c^2$ ($1.8\times10^{52}\,\mathrm{erg}$) in GWs. 
We see that in each case our exclusion distances are of order 100\,kpc (the analyses of GRB\,200412A and 
GRB\,200420A did not produce 90\% exclusion distances for the 1000\,Hz sine--Gaussians above 10\,kpc).  
This is not enough to test the magnetar giant flare hypothesis for GRB\,200415A \cite{Svinkin:2021wcp}.

\begin{figure}
\centering
\includegraphics[width=0.7\textwidth]{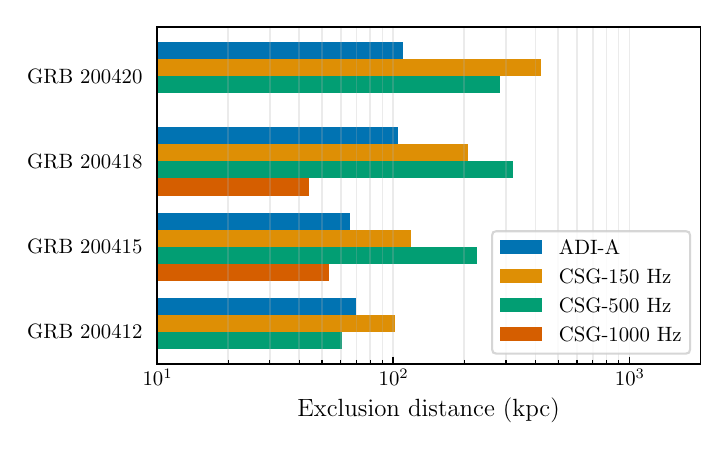}
\caption{
    The 90\% confidence-level exclusion distances for each of the \acp{GRB} analysed by the \xpipeline generic burst 
search, for the accretion disk instability (ADI) signal model A
and for circular sine--Gaussian (CSG) signals at 150\,Hz, 500\,Hz, and 1000\,Hz. 
    For a given \ac{GRB} and signal model this is the distance within which 90\% of simulated signals inserted 
into off-source data are recovered passing all background rejection tests and with a significance greater than the loudest on-source 
candidate event (if any).
\label{fig:xpipeline}}
\end{figure}

\section{Summary and Discussion}
\label{sec:conclusions}

We have presented the results of the first joint observation of the KAGRA detector with GEO, 
performed during April 7--20 2020.  The coincident observational data from GEO and KAGRA were analysed jointly 
to look for transient \ac{GW} signals, including neutron-star binary coalescences and generic unmodeled transients. 
We also performed dedicated searches for \ac{CBC} signals and generic transients associated with 
\acp{GRB} observed during the joint run. 
No candidate \ac{GW} events were identified.

In the all-sky \ac{BNS} search, the most significant candidate from the analysis of 0.032 years of data 
has an \ac{iFAR} of 0.033\,years, consistent with background.
The sensitive spacetime volume to \ac{BNS} coalescences was estimated as a function of \ac{iFAR},
and we found that the \ac{iFAR} of the most significant event corresponds to a sensitive distance 
of $\sim$0.6\,Mpc, comparable to that expected from the noise spectra. 

In the all-sky burst search, the most significant candidate from the analysis of 0.012 years of data 
has an \ac{iFAR} of $0.097$ years which is not significant enough to be considered a likely \ac{GW} event. 
The sensitivity of the search was estimated in terms of the minimal detectable root-sum-square signal amplitude 
and minimum detectable signal energy at a fixed distance. 
We find minimal detectable energies of around $10^{-6}\,M_\odot c^2$ to $10^{-3}\,M_\odot c^2$ for sources at 10\,kpc.
These sensitivities are consistent with the amplitude spectral densities of the detectors.

The searches for \acp{CBC} and generic transient signals associated with \acp{GRB} found no candidate events, 
with the lowest $p$-value for any \ac{GRB} being \pvalBurstLowest.
For GRB\,200415A, the dedicated \ac{CBC} search set a 90\% exclusion distance of \DBNSAprFifteen for \ac{BNS} systems,
\DNSBHGenAprFifteen for generically spinning NSBH, and \DNSBHAliAprFifteen for aligned-spin NSBH. 
At a distance of \mbox{3.5\,Mpc}, corresponding to NGC 253,
the exclusion confidences for these populations 
are \pcBNSAprFifteenNGC, \pcNSBHGenAprFifteenNGC and \pcNSBHAliAprFifteenNGC respectively. 
The sensitivity of the generic burst search was evaluated for several GW emission models, 
giving 90\% exclusion distances of order 100\,kpc for sources emitting $10^{-2}\,M_\odot c^2$ energy in GWs. 
These results are not strong enough to test the binary merger or magnetar hypotheses for the progenitor of GRB\,200415A.

The lack of detected \acp{GW} in this run is expected given the sensitivity of the GEO--KAGRA network at the time. 
However, the sensitivity of KAGRA is expected to improve by more than two orders of magnitude 
later in this decade \cite{KAGRA:2013pob}, becoming comparable to that of the LIGO and Virgo detectors.
Our analyses have demonstrated the ability to incorporate KAGRA data into standard transient search pipelines 
that have been used to detect \acp{GW} in LIGO and Virgo data.
Adding KAGRA to the LIGO--Virgo network will improve the sky-localization accuracy
and increase the number of events detected with 3 or more detectors simultaneously
\cite{Wen:2010cr,Schutz:2011tw}.
KAGRA is planning to join the forth observing run of the advanced-detector network. 
We look forward to KAGRA's scientific contributions in the coming years as a member of the global \ac{GW} detector network.

The full O3GK detector strain data and data products associated with this paper are available through Gravitational Wave Open Science Center \cite{O3GKDataRelease}.

\section*{Acknowledgments}
This material is based upon work supported by NSF’s LIGO Laboratory which is a major facility
fully funded by the National Science Foundation.
The authors also gratefully acknowledge the support of
the Science and Technology Facilities Council (STFC) of the
United Kingdom, the Max-Planck-Society (MPS), and the State of
Niedersachsen/Germany for support of the construction of Advanced LIGO 
and construction and operation of the GEO\,600 detector. 
Additional support for Advanced LIGO was provided by the Australian Research Council.
The authors gratefully acknowledge the Italian Istituto Nazionale di Fisica Nucleare (INFN),  
the French Centre National de la Recherche Scientifique (CNRS) and
the Netherlands Organization for Scientific Research (NWO), 
for the construction and operation of the Virgo detector
and the creation and support  of the EGO consortium. 
The authors also gratefully acknowledge research support from these agencies as well as by 
the Council of Scientific and Industrial Research of India, 
the Department of Science and Technology, India,
the Science \& Engineering Research Board (SERB), India,
the Ministry of Human Resource Development, India,
the Spanish Agencia Estatal de Investigaci\'on (AEI),
the Spanish Ministerio de Ciencia e Innovaci\'on and Ministerio de Universidades,
the Conselleria de Fons Europeus, Universitat i Cultura and the Direcci\'o General de Pol\'{\i}tica Universitaria i Recerca del Govern de les Illes Balears,
the Conselleria d'Innovaci\'o, Universitats, Ci\`encia i Societat Digital de la Generalitat Valenciana and
the CERCA Programme Generalitat de Catalunya, Spain,
the National Science Centre of Poland and the European Union -- European Regional Development Fund; Foundation for Polish Science (FNP),
the Swiss National Science Foundation (SNSF),
the Russian Foundation for Basic Research, 
the Russian Science Foundation,
the European Commission,
the European Social Funds (ESF),
the European Regional Development Funds (ERDF),
the Royal Society, 
the Scottish Funding Council, 
the Scottish Universities Physics Alliance, 
the Hungarian Scientific Research Fund (OTKA),
the French Lyon Institute of Origins (LIO),
the Belgian Fonds de la Recherche Scientifique (FRS-FNRS), 
Actions de Recherche Concertées (ARC) and
Fonds Wetenschappelijk Onderzoek – Vlaanderen (FWO), Belgium,
the Paris \^{I}le-de-France Region, 
the National Research, Development and Innovation Office Hungary (NKFIH), 
the National Research Foundation of Korea,
the Natural Science and Engineering Research Council Canada,
Canadian Foundation for Innovation (CFI),
the Brazilian Ministry of Science, Technology, and Innovations,
the International Center for Theoretical Physics South American Institute for Fundamental Research (ICTP-SAIFR), 
the Research Grants Council of Hong Kong,
the National Natural Science Foundation of China (NSFC),
the Leverhulme Trust, 
the Research Corporation, 
the Ministry of Science and Technology (MOST), Taiwan,
the United States Department of Energy,
and
the Kavli Foundation.
The authors gratefully acknowledge the support of the NSF, STFC, INFN and CNRS for provision of computational resources.
This work was supported by MEXT, JSPS Leading-edge Research Infrastructure Program, JSPS Grant-in-Aid for Specially Promoted Research 26000005, JSPS Grant-in-Aid for Scientific Research on Innovative Areas 2905: JP17H06358, JP17H06361 and JP17H06364, JSPS Core-to-Core Program A. Advanced Research Networks, JSPS Grant-in-Aid for Scientific Research (S) 17H06133 and 20H05639 , JSPS Grant-in-Aid for Transformative Research Areas (A) 20A203: JP20H05854, the joint research program of the Institute for Cosmic Ray Research, University of Tokyo, National Research Foundation (NRF), Computing Infrastructure Project of KISTI-GSDC, Korea Astronomy and Space Science Institute (KASI), and Ministry of Science and ICT (MSIT) in Korea, Academia Sinica (AS), AS Grid Center (ASGC) and the Ministry of Science and Technology (MoST) in Taiwan under grants including AS-CDA-105-M06, Advanced Technology Center (ATC) of NAOJ, and Mechanical Engineering Center of KEK.
{\it We would like to thank all of the essential workers who put their health at risk during the COVID-19 pandemic, without whom we would not have been able to complete this work.}

\let\doi\relax

\bibliographystyle{ptephy}

\end{document}